\documentclass[12pt]{article}   
\usepackage{fullpage}
\usepackage[super,sort,comma]{natbib}

\usepackage{fancyhdr}		

\usepackage{amsmath,amssymb,amsfonts}
\usepackage{algorithm}
\usepackage{algpseudocode}
\usepackage{hyperref}
\usepackage{graphicx}
\usepackage{multirow}
\usepackage{subfig}
\usepackage{makecell}

\title{Equivariant Multiscale Learned Invertible Reconstruction for Cone Beam CT}
\author{Nikita Moriakov \and Jan-Jakob Sonke \and Jonas Teuwen \\Department of Radiation Oncology, Netherlands Cancer Institute, \\the Netherlands}

\begin{document}
\maketitle

\pagenumbering{roman}
\setcounter{page}{1}
\pagestyle{plain}

\begin{abstract}
  \noindent {\bf Background:} Cone Beam CT (CBCT) is an essential imaging modality nowadays, but the image quality of CBCT still lags behind the high quality standards established by the conventional Computed Tomography (CT). Deep learning reconstruction methods have emerged as a promising alternative to classical iterative reconstruction methods, but applying such methods to CBCT is often difficult due to memory limitations and the need for fast inference at clinically-relevant resolutions. \\ 
{\bf Purpose:} We propose LIRE+, a learned iterative scheme for fast and memory-efficient CBCT reconstruction, which is a substantially faster and more parameter-efficient alternative to the recently proposed LIRE method. \\
{\bf Methods:} LIRE+ is a rotationally-equivariant multiscale learned invertible primal-dual iterative scheme for CBCT reconstruction. Memory usage is optimized by relying on simple reversible residual networks in primal/dual cells and patch-wise computations inside the cells during forward and backward passes, while increased inference speed is achieved by making the primal-dual scheme multiscale so that the reconstruction process starts at low resolution and with low resolution primal/dual latent vectors. Transitions to higher resolutions are performed with nearest upsampling operations, whose injectivity allows to reverse the operation unambiguously during the backward pass. Rotational equivariance is accomplished with group equivariant convolutions inside the primal cells.\\
{\bf Results:} A LIRE+ model was trained and validated on a set of 260 + 22 thorax CT scans and tested using a set of 142 thorax CT scans with additional evaluation with and without finetuning on an out-of-distribution set of 79 Head and Neck (HN) CT scans. Our method surpasses classical and deep learning baselines, including LIRE, on the thorax test set. For a similar inference time and with only 37 \% of the parameter budget, LIRE+ achieves a +0.2 dB Peak Signal-to-Noise Ratio (PSNR) improvement over LIRE, while being able to match the performance of LIRE in 45 \% less inference time and with 28 \% of the parameter budget. Rotational equivariance ensures robustness of LIRE+ to patient orientation, while LIRE and other deep learning baselines suffer from substantial performance degradation when patient orientation is unusual. On the HN dataset in the absence of finetuning, LIRE+ is generally comparable to LIRE in performance apart from a few outlier cases, whereas after identical finetuning LIRE+ demonstrates a +1.02 dB PSNR improvement over LIRE. \\
{\bf Conclusions:} Multiscale reconstruction can be naturally integrated into invertible learned primal-dual scheme and can accelerate CBCT reconstruction without loss of image quality. Rotational equivariance in a learned primal-dual iterative scheme can be enforced by making the primal components of the network rotationally equivariant, improving robustness of the network to unusual patient orientation.\\

\end{abstract}

\tableofcontents

\newpage

\pagenumbering{arabic}
\setcounter{page}{1}
\pagestyle{fancy}
\section{Introduction}
\label{s.intro}
Computed Tomography (CT) is one of the most used medical imaging modalities nowadays. Similar to many other modern imaging modalities such as MRI, the measurements acquired by a CT scanner - i.e., X-ray projection images taken from a multitude of angles - are not immediately usable in clinic and instead need to undergo the process of \emph{reconstruction}, wherein they are processed by a reconstruction algorithm and combined into a three-dimensional volume. An important variation of CT is Cone Beam Computed Tomography (CBCT), where the X-ray source emits rays in a wide cone-shaped beam and the detector is a large flat panel array. In CBCT, both the X-ray source and the detector typically follow circular trajectories around the isocenter, and the detector is sometimes offset to give a larger field of view\cite{letourneau2005}. CBCT has applications in interventional radiology\cite{floridi2014}, dentistry\cite{dawood2009} and image-guided radiation therapy\cite{jaffray2002}, however, CBCT image quality remains poor compared to classical CT with helical trajectory for a few reasons. CBCT reconstruction is inherently harder since the data completeness condition for exact reconstruction of the whole volume is not satisfied for circular source/detector orbits \cite{maas2010, tuy1983}. Photon starvation, particularly in highly attenuated areas and in lower-dose scans, results in strong streaking artifacts. Scattering becomes a bigger issue as well, since a large detector panel captures more scattered photons from a wide cone beam of X-rays. Resulting poor Hounsfield Unit (HU) calibration is a limitating factor for applications in e.g. adaptive radiotherapy, where a daily CBCT scan for treatment plan adjustment without the need for registration to a prior CT scan\cite{sonke2019} would be very desirable.

Deep learning reconstruction methods have drawn a lot interest from the medical imaging community by achieving remarkable results in public reconstruction challenges such as FastMRI\cite{fastmri2020, beauferris2020}. Reconstruction methods in the \emph{learned post-processing} family apply a neural network as a learned operator on top of a classical reconstruction method such as filtered back-projection (FBP)\cite{unser2017, zhi2021}. Despite the advantages such as typically fast inference, learned post-processing methods do not provide the neural network with direct access to the underlying measurement data, thus some imaging artifacts might be hard to fix. For example, removing streaks due to photon starvation in CBCT in image domain would require a neural network with large receptive field due to the size of the streaks. \textit{Learned iterative schemes}, on the other hand, are inspired by classical iterative methods such as Landweber iteration, and embed the forward operator directly in the neural network architecture. Intuitively, this allows to draw on the theoretical guarantees given by iterative methods but use more flexible `neural network prior' instead of an explicit regularization. Learned Primal-Dual (LPD) algorithm\cite{Adler2017b} is a prominent example of a learned iterative scheme inspired by the Primal-Dual Hybrid Gradient (PDHG) method\cite{pdhg2011}, which combines both image-space and projection-space operations in an end-to-end trainable network. Image-space computations are performed by \emph{primal blocks} and projection-space computations are performed by \emph{dual blocks}, all primal/blocks being small convolutional neural networks. LPD framework has been extended to other modalities as well, such as Digital Breast Tomosynthesis\cite{teuwen2021} and MRI\cite{ramzi2020a}, but there are also recent examples of learned iterative schemes for CT\cite{airnet} or MRI\cite{yiasemis2022} reconstruction that work in image domain only.

However, learned iterative schemes and LPD in particular can be hard to scale up to a fully three-dimensional modality such as CBCT due to memory limitations. For example,  given a $256 \times 256 \times 256$ FP32 tensor a \emph{single} convolution layer with $64$ features would already require 8 GB memory to perform the backpropagation operation. One of the first memory-efficient alternatives is $\partial $U-Net\cite{hauptmann2020}, which is a simpler scheme that does not operate in the projection space. Memory usage is reduced by relying on a multiscale approach, where reconstructions obtained at different resolutions are merged together by a final U-net. iLPD, or invertible learned primal-dual method, has been considered\cite{rudzusika2021}, where it was shown that it substantially reduces memory requirements and allows to use longer learned iterative schemes. For a 3D helical CT setting, iLPD has been combined\cite{bajic2022} with splitting the scanning geometry in chunks of data that can be processed independently, however, such geometry splitting is not possible for CBCT. To address this issue, LIRE\cite{lire2023} method was recently proposed, where a learned invertible primal-dual scheme was augmented with tiling computation mechanism inside the primal/dual blocks during both training and inference, allowing to use higher filter counts as well more complex U-net cells inside primal blocks. LIRE inference takes around 30 seconds on NVIDIA A100 accelerator with clinically relevant geometry and resolution, and it is desirable to speed it up for future clinical application. A logical step would be to try to combine learned invertible primal-dual scheme and multiscale reconstruction\footnote{It might appear counterintuitive, since the input and the output in a reversible neural network have the same dimensionality, but will be explained in Section \ref{ss.lirep}.}, but it has not been done in literature at the moment, even though invertible flows that incorporate multiscale latent codes are well known in generative modeling\cite{glow2018}.

Parallel to the development of new learned iterative schemes for reconstruction, the study of natural symmetries of learning tasks and the means of building these symmetries into neural network architectures has been a fruitful recent research direction in inverse problems and deep learning in general. For instance, when a patient is rotated we expect the new reconstruction to be a rotated version of the original reconstruction. For convolutional neural networks, this problem is addressed with group equivariant convolutions\cite{cohen2016}, which often allow to achieve state of the art results at reduced parameter cost on image classification tasks. Group equivariant convolutional neural networks have been applied to inverse problems with learned iterative schemes as well\cite{celledoni2021}, but not in the context of CBCT and learned-primal dual family of methods. Additionally, even though group equivariant convolutional neural networks allow to reduce the parameter cost, the dimensionality of internal representations is typically increased\footnote{This is the consequence of a rule of thumb\cite{cohen2016}, where the number of filters is reduced by the square root of the group size to approximately match the total parameter count.} resulting in increased inference times. Therefore, including group-equivariant operations in CBCT reconstruction further necessitates the search for fast and memory efficient learned iterative schemes.

In this work, we address these research questions and improve upon the LIRE method for CBCT reconstruction by designing LIRE+, a faster and more parameter-efficient learned primal-dual scheme yielding similar or better reconstruction quality, which additionally enjoys rotational equivariance for improved robustness to unusual patient orientation. We perform extensive evaluation of LIRE+ and the baselines using image quality metrics such as PSNR and Structural Similarity Index Measure (SSIM), as well as HU Mean Absolute Error (MAE).

\section{Methods}
\label{s.matmethods}
In order to have a direct comparison for the novel method on the same dataset, we follow the experimental setup of LIRE\cite{lire2023}, which we briefly recollect in Sections \ref{ss.tomo}, \ref{ss.data} and \ref{ss.baselines}.

\subsection{Tomography}
\label{ss.tomo}
The \textit{cone-beam transform operator}, or simply the \textit{projection operator}, is defined as an integral operator
\begin{equation}
\mathcal P(x)(t, u) = \int_{L_{t, u}} x(z) dz, 
\end{equation}
where $x: \Omega_X \to \mathbb R$ is a function specifying attenuation coefficients in the spatial domain $\Omega_X \subset \mathbb R^3$ and $L_{t, u}$ is a line from the source to the detector element $u$ at time $t$. $\mathcal P$ is a linear operator, and Hermitian\footnote{For suitably defined $L^2$ function spaces.} adjoint $\mathcal P^*$ of $\mathcal P$ is called the \textit{backprojection operator}. Using the projection operator $\mathcal P$, we model noisy CBCT acquisition as
\begin{equation}\label{eq.noisemodel}
y = \text{\texttt{Poisson}}(I_0 \cdot e^{-\mathcal P x}),
\end{equation}
where $I_0$ is the unattenuated X-ray photon count. The inverse problem of CBCT reconstruction is then to determine the tissue attenuation coefficients $x$ from the noisy projection data $y$.

We will approach the inverse problem in \eqref{eq.noisemodel} by finding a Bayes estimator parametrized by a neural network. The goal for the Bayes estimator $\hat x_{\textrm{Bayes}}$ in general is to minimize the expected cost
\begin{equation}\label{eq.bayesloss}
L(\hat x) = \mathbb E_{(x, y) \sim \pi} \ L(x, \hat x(y))
\end{equation}
over all estimators $\hat x$, where $\pi$ is the distribution of pairs $(x, y)$ of tomographic volumes $x$ with the corresponding projection images $y$, and $L$ is a fixed cost function. In this work, a sum of mean absolute error and a Structural Similarity loss will play the role of the cost function $L$, the optimal estimator in \eqref{eq.bayesloss} will be chosen from a certain class of neural networks, and minimization of the cost in \eqref{eq.bayesloss} with respect to the parameters of the network will be carried out via minibatch stochastic gradient descent. We refer to Section 5.1.2\cite{Kaipio2005} for more information on Bayes estimators.

\subsection{Data}
\label{ss.data}
In this work we simulate a common clinical acquisition geometry for a  Linac-integrated CBCT scanner from Elekta\cite{letourneau2005} with a medium field-of-view setting, offset detector, a full $2 \pi$ scanning trajectory and $720$ projections. The source-isocenter distance is $1000$ mm and the isocenter-detector plane distance is $536$ mm. The detector is offset by $115$ mm to the side in the direction of rotation to give an increased Field of View. Square detector panel with a side of $409.6$ mm and $256 \times 256$ pixel array was used.

To train and evaluate our model, we used the same thorax CT and head \& neck CT data as LIRE\cite{lire2023}, consisting of a dataset of 424 thorax CT scans with isotropic spacing of $1$ mm and a dataset of 79 head \& neck CT scans with anisotropic spacing of between $0.9$ mm and $1.0$ mm for axial plane and between $1.0$ mm and $1.6$ mm for the perpendicular direction. Both datasets had axial slice of $512 \times 512$ voxels. All data was downsampled to $2$ mm isotropic resolution, resulting in volumes with $256^3$ voxels. No denoising was applied to the CT scans, since unsupervised denoising could blur very fine details such as fissures leading to over-optimistic image quality metrics. Study approval was granted by the IRB of our institute, IRBd20-008.

The thorax CT dataset was split into a training set of 260 scans, a validation set of 22 scans and a test set of 142 scans. The additional head \& neck dataset was used in two regimes: for testing the models on out-of-distribution data and for additional finetuning experiment, where 79 volumes were randomly partitioned into a finetuning set of 8 volumes, 2 volumes for validation and 69 for testing\footnote{Four outlier head \& neck cases, discussed in Section \ref{s.results}, were used as test data.}. To simulate noisy projection data from the CT scans, Hounsfield units were converted into attenuation coefficients using $\mu = 0.2 \ \textrm{cm}^{-1}$ as the water linear attenuation coefficient. Attenuated projection data was corrupted by Poisson noise with $I_0 = 30000$ photons in \eqref{eq.noisemodel}.

\subsection{Baseline methods}
\label{ss.baselines}
We provide a direct comparison with LIRE given that we are using the same dataset, and to add context to this comparison we will cite the following baselines from the original work\cite{lire2023}: FBP\cite{feldkamp84}, PDHG\cite{pdhg2011} with Total Variation (TV) regularisation, U-net\cite{cicek2016} and Uformer\cite{wang2022,yang2023,chen2022} with FBP input, as well as LIRE itself. A new additional baseline is $\partial$U-Net\cite{hauptmann2020}, which plays the role of an alternative memory efficient and fast learned iterative scheme. Our implementation of $\partial$U-Net relies on the open-source implementation\footnote{Adapted to 3D and our projector/backprojector code from \url{https://github.com/asHauptmann/multiscale}} from the author, where the base filter count was increased from 12 to 32 in order to get closer to the filter counts used by LIRE and LIRE+ to make the comparison fair but fit into memory budget. We switched from batch normalization to instance normalization in our version of $\partial$U-Net, since batch normalization resulted in unstable convergence, presumably, because of very small batch size. As input to $\partial$U-net, we provided the FBP reconstruction and the field-of-view tensor $V$ defined later in Section \ref{ss.lirep}. The same augmentation strategy as LIRE+ and the same loss function (see Section \ref{ss.lirep}) were used. To train $\partial$U-net, Adam optimizer\cite{kingma2014} was employed with batch size of $8$ on two NVIDIA Quadro RTX 8000 cards via gradient accumulation, initial learning rate of $0.0001$ and a plateau scheduler with linear warm-up and 10 epoch patience. The best-performing model on the validation set was chosen for testing. For the finetuning experiment on HN data, the corresponding pretrained $\partial$U-net or LIRE model was taken and finetuned for 50 epochs with a quarter of the initial learning rate using a combined dataset of 8 HN CT volumes and 8 randomly chosen thorax CT volumes; best performing model on HN validation set was picked for the final testing on HN test data.

\subsection{LIRE+}
\label{ss.lirep}

\label{ss.arch}
\begin{algorithm}
\centering
\begin{algorithmic}[1]
\Procedure{\texttt{reconstruct}}{$y, \mathcal P, \mathcal P^*, \theta, V$}
\State $\alpha \gets 50\%$ \Comment{Set resolution to 50\%}
\State $\overline y \gets \text{\texttt{ProjDown}}(y)$ \Comment{Downsample \& subsample projections}
\State $\overline x \gets \mathcal P_{\alpha}^*(\overline y)$ \Comment{Normalized backprojection initialization}
\State $\overline V \gets \text{\texttt{Downsample}}(V)$ \Comment{Downsample FoV tensor}
\State $I \gets []$ \Comment{Initialize output list}
\State $f \gets \overline x^{\otimes 8} \in X^{8}$\Comment{Initialize primal vector}
\State $h \gets \overline y^{\otimes 8} \in U^{8}$\Comment{Initialize dual vector}
\For{$i \gets 1, \dots, 12$}
\State $d_1, d_2 \gets \text{\texttt{Splt}}(h)$ \Comment{Split dual channels}
\State $p_1, p_2 \gets \text{\texttt{Splt}}(f)$ \Comment{Split prime channels}
\State $p_{\text{op}} \gets \mathcal P_{\alpha}([p_2, \overline x]^{\oplus})$ \Comment{Project $p_2$ and $\overline x$}
\State $d_2 \gets d_2 + \Gamma_{\theta_i^d}([p_{\text{op}}, d_1, \overline y]^{\oplus})$ \Comment{Upd. $d_2$}
\State $b_{\text{op}} \gets \mathcal P_{\alpha}^*(d_2)$ \Comment{Backproject $d_2$}
\State $\text{\textit{LW}} \gets \mathcal P_{\alpha}^* (\mathcal P_{\alpha} (\overline x) - \overline y)$ \Comment{Landweber term}
\State $p_2 \gets p_2 + \Lambda_{\theta_i^p}([b_{\text{op}}, p_1, \overline x, \text{\textit{LW}}, \overline V]^{\oplus})$ \Comment{Upd. $p_2$}
\State $h \gets [d_1, d_2]^{\oplus}$ \Comment{Combine new dual}
\State $f \gets [p_1, p_2]^{\oplus}$ \Comment{Combine new primal}
\State $\overline x \gets \overline x + \text{\texttt{Conv3d}}(f, \theta_i^o)$ \Comment{Update reconstruction}
\State $I \gets I + [\overline x]$ \Comment{Append $\overline x$ to output list}
\State $h \gets \text{\texttt{Perm}}(h, {\theta_i^m})$ \Comment{Permute dual channels w. $\theta_i^m$}
\State $f \gets \text{\texttt{Perm}}(f, {\theta_i^m})$ \Comment{Permute prim. channels w. $\theta_i^m$}
\If{$i == 6$}
\State $f, h \gets \text{\texttt{Upsample}}(f), \text{\texttt{Upsample}}(h)$ \Comment{Upsample latent vectors}
\State $\overline x \gets \text{\texttt{Upsample}}(\overline x)$ \Comment{Upsample reconstruction}
\State $\overline y, \overline V, \alpha \gets y, V, 100\%$ \Comment{Switch to full resolution}
\EndIf
\EndFor
\State \textbf{return} $I$
\EndProcedure
\end{algorithmic}
\caption{LIRE+}
\label{alg:liremain}
\end{algorithm}

LIRE+ method is an unrolled learned iterative scheme, which extends LIRE by relying on multiscale reconstruction strategy to improve the inference speed, equivariant primal cells for higher parameter efficiency and robustness to orientation, as well as centered weight normalization to improve convergence stability. Similar to LIRE, the memory footprint of LIRE+ is reduced by combining invertibility for the network as a whole and patch-wise computations for local operations. An optional CPU-GPU memory streaming mechanism is implemented, which would keep entire primal/dual vectors in CPU memory and only send the patch required for computing the primal/dual updates or gradients into the GPU. We refer the reader to the original work\cite{lire2023} for the discussion on invertibility and patch-wise computations. To justify the combination of multiscale reconstruction and invertibility, we make the following observation: if $\Lambda: \mathbb R^n \to \mathbb R^n$ is an invertible neural network and $\iota: \mathbb R^n \to \mathbb R^m, m \geq n$ is some fixed injective differentiable mapping such as nearest upsampling operation, then the input $x \in \mathbb R^n$ can be restored from the output $\iota(\Lambda(x)) \in \mathbb R^m$ unambigously by first inverting $\iota$ and then $\Lambda$, so the gradients for the parameters of $\Lambda$ can be computed without storing the activations during the forward pass. The algorithm was implemented as a C++/CUDA extension for PyTorch\cite{pytorch} in order to maximize memory efficiency, training and inference speed.

LIRE+, given by function \texttt{RECONSTRUCT}($y, \mathcal P, \mathcal P^*, \theta, V$) in Algorithm~\ref{alg:liremain}, consists of 12 iterations and uses primal/dual latent vectors with 8 channels. Here $y$ is log-transformed and scaled projection data, $\mathcal P$ and $\mathcal P^*$ are normalized projection and backprojection operators respectively, $\theta$ is a list of parameters and $V$ is an auxiliary Field-of-View tensor defined as
\begin{equation*}
V(p) = \begin{cases} 
      1 & p \textrm{ is seen from all projection angles} \\
      0.5 & p \textrm{ is seen from half of the proj. angles} \\
      0 & \textrm{otherwise.}
   \end{cases}
 \end{equation*}
The parameters $\theta$ are partitioned into 4 parameter groups, where $\{ \theta_i^p \}_{i=1}^{12}$ are the primal block parameters, $\{ \theta_i^d \}_{i=1}^{12}$ are the dual block parameters, $\{ \theta_i^o \}_{i=1}^{12}$ are the output convolution parameters and $\{ \theta_i^m \}_{i=1}^{12}$ are the permutation parameters. For every $i$, the permutation $\theta_i^m$ is some fixed permutation of $[1,2,\dots,8]$ which is randomly initialized during model initialization and stored as a model parameter; we require that $\theta_i^m$ mixes the first and the second half of $[1,2,\dots,8]$. Channel-wise concatenation of tensors $z_1, z_2, \dots, z_k$ is denoted by $[z_1, z_2, \dots, z_k]^{\oplus}$, conversely, function $\text{\texttt{Splt}}(z)$ splits tensor $z$ with $2n$ channels into two halves along the channel dimension. Function $\text{\texttt{Perm}}(z, \rho)$ permutes tensor $z$ with $n$ channels along the channel dimension with the permutation $\rho \in \text{\texttt{Sym}}(n)$. Function $\text{\texttt{Upsample}}(z)$ performs nearest upsampling of $z$ to twice the resolution, $\text{\texttt{Downsample}}(z)$ downsamples tensor $z$ to half the resolution with linear interpolation and function $\text{\texttt{ProjDown}}(z)$ downsamples projection tensor $z$ to half the resolution and drops every second projection. For resolution $\alpha \in [50\%, 100\%]$, we write $\mathcal P_{\alpha}, \mathcal P_{\alpha}^*$ for the projection and backprojection operator respectively at $\alpha$ resolution, where for half resolution only every second projection from the original data is computed.
 
LIRE+ uses a number of convolutional blocks. $\text{\texttt{Conv3d}}(\cdot, \theta_i^o)$ denotes a $1 \times 1 \times 1$ convolution with parameters $\theta_i^o$. $\Gamma_{\theta_i^d}$ denotes $i$-th dual block with parameters $\theta_i^d$ comprised of 3 layers of $3 \times 3 \times 3$ convolutions with 96, 96 and 4 filters respectively and LeakyReLU activation after the first and the second convolution layers. $i$-th primal block with parameters $\theta_i^p$ is denoted by $\Lambda_{\theta_i^p}$, which is comprised of 3 layers of $3 \times 3 \times 3$ P4-equivariant convolutions with 64, 64 and 4 filters respectively and LeakyReLU activation after the first and the second convolution layers. That is, input to a primal block has $8$ channels and no `group dimension', the first and the second convolution layers in a primal block return tensors with $64 \times 4$ channels (number of filters $\times$ group size), while output of the last convolution has $4 \times 4$ channels. This output is then averaged over the group dimension, making the primal block equivariant w.r.t. the action of P4 (i.e., $90$-degree rotations along the z-axis). We used centered weight normalization\cite{huang2017} for the primal/dual block parameters to improve training stability.

LIRE+ training procedure is identical to LIRE, importantly, the loss function is a weighted sum of $L^1$ norm $\| \cdot \|$ and SSIM loss, taken separately over the full field of view region (i.e., voxels present in at least half of the projections) and the partial field of view region (i.e., voxels present in at least one projection). Mathematically,
\begin{align*}
L(x, y) &= \| x - y \|_{\text{FullFoV}} + \alpha_1 \cdot (1.0 - \text{\texttt{SSIM}}_{\text{FullFoV}}(x, y)) + \\
  & + \alpha_2 \cdot \| x- y \|_{\text{PartFoV}} + \alpha_2 \cdot \alpha_1 \cdot (1.0 - \text{\texttt{SSIM}}_{\text{PartFoV}}(x, y)),
\end{align*}
where $\alpha_1 = 0.5$ and $\alpha_2$ was set to $0.1$ initially and then reduced to $0.01$ after first learning rate decay step. The algorithm returns a list $I = [x_1, x_2, \dots, x_{12}]$ of reconstructions where the first $6$ elements have half the resolution and the last $6$ elements have full resolution. Reconstruction losses for all $x \in I$ are computed and summed, the ground truth is downsampled to compute the loss for the half-resolution reconstructions $x_1,\dots,x_6$. As a data augmentation strategy, we randomply flipped along the left-right and the head-foot axes. Isocenter was chosen by adding a random offset sampled from an isotropic Gaussian distribution with $0$ mm mean and a standard deviation of $100$ mm to the volume center. 

LIRE+ was trained to reconstruct complete volumes. Two NVIDIA A100 GPUs with gradient accumulation were used with gradient accumulation enabled to achieve effective batch size of $8$. Adam optimizer\cite{kingma2014} was employed with an initial learning rate of $0.001$ and a plateau scheduler with linear warm-up and 10 epoch patience. At the end of each epoch models were evaluated, the best model was picked for testing. For the finetuning experiment on HN data, LIRE+ was finetuned for 50 epochs with a quarter of the initial learning rate using a combined dataset of 8 HN CT volumes and 8 randomly chosen thorax CT volumes; best performing model on HN validation set was picked for the final testing on HN test data.

\section{Results}
\label{s.results}
\subsection{Memory usage, parameter count, inference speeds}
For the internal patch-based computations inside LIRE+ we set the patch size to $128 \times 128 \times 128$, resulting in roughly $30$ GB VRAM usage per single volume during training. Reducing the patch size to $32 \times 32 \times 32$ and enabling CPU-GPU streaming decreased the usage to roughly $12$ GB VRAM per single volume. For $\partial$U-net, GPU memory usage during training was around $48$ GB per volume.

The total parameter counts are: 24M parameters for LIRE, 7M for LIRE+ with 9 iterations, 9M for LIRE+ with 12 iterations and 27M for $\partial$U-net. We measured the following per-volume inference times on NVIDIA A100 accelerator: 31 seconds for LIRE, 32 seconds for LIRE+ with 12 iterations, 17 seconds for LIRE+ with 9 iterations and 4 seconds for $\partial$U-net.  

\subsection{Image quality: thorax CT}
\begin{table*}[h]
\caption{Test results on thorax CT (best result in bold), mean $\pm$ std.dev.}
\label{tab:str-llung}
\centering
\begin{tabular}{|l|l l l l|}
\hline
Method & PSNR & SSIM & MAE 2mm (HU)  & MAE 4mm (HU) \\
\hline
FBP & $20.05 \pm 2.30$ & $0.66 \pm 0.07$ & $270.70 \pm 19.78$ & $261.29 \pm 21.36$ \\
TV & $29.23 \pm 2.87$ & $0.79 \pm 0.09$ & $85.07 \pm 24.10$ & $35.26 \pm 8.32$ \\
Uformer & $31.62 \pm 2.44$ & $0.81 \pm 0.06$ & $62.91 \pm 7.44$ & $43.66 \pm 5.74$ \\
U-Net & $34.29 \pm 2.71$ & $0.84 \pm 0.06$ & $47.86 \pm 7.34$ & $21.70 \pm 3.56$ \\
$\partial$U-Net & $34.55 \pm 2.72$ & $0.90 \pm 0.05$ & $46.14 \pm 7.01$ & $15.71 \pm 2.71$\\
LIRE & $35.14 \pm 2.76$ & $0.91 \pm 0.04$ & $43.02 \pm 6.88$ & $13.20 \pm 2.64$ \\
LIRE+ 9 it. & $35.15 \pm 2.79$ & $0.91 \pm 0.05$ & $42.89 \pm 6.99$ & $13.58 \pm 2.66$ \\
LIRE+ 12 it. & $\mathbf{35.38 \pm 2.82}$ & $\mathbf{0.91 \pm 0.04}$ & $\mathbf{41.86 \pm 7.01}$ & $\mathbf{13.11 \pm 2.64}$ \\
\hline
\end{tabular}
\end{table*}

\begin{table*}[h]
\caption{Test results on rotated thorax CT (best result in bold), mean $\pm$ std.dev.}
\label{tab:r90-llung}
\centering
\begin{tabular}{|l|l l l l|}
\hline
Method & PSNR & SSIM & MAE 2mm (HU)  & MAE 4mm (HU) \\
\hline
Uformer & $29.53 \pm 2.58$ & $0.80 \pm 0.06$ & $70.70 \pm 8.83$ & $49.06 \pm 6.64$ \\
U-Net & $29.98 \pm 2.62$ & $0.83 \pm 0.06$ & $60.48 \pm 8.55$ & $27.99 \pm 3.73$ \\
$\partial$U-Net & $33.64 \pm 2.66$ & $0.89 \pm 0.05$ & $49.65 \pm 6.95$ & $17.50 \pm 2.82$\\
LIRE & $34.71 \pm 2.72$ & $0.91 \pm 0.05$ & $44.60 \pm 6.85$ & $14.17 \pm 2.78$ \\
LIRE+ 9 it. & $35.15 \pm 2.79$ & $0.91 \pm 0.05$ & $42.89 \pm 6.99$ & $13.58 \pm 2.66$ \\
LIRE+ 12 it. & $\mathbf{35.38 \pm 2.82}$ & $\mathbf{0.91 \pm 0.04}$ & $\mathbf{41.86 \pm 7.01}$ & $\mathbf{13.12 \pm 2.64}$ \\
\hline
\end{tabular}
\end{table*}

We perform extensive evaluation of LIRE+ and the baselines using image quality metrics such as PSNR and SSIM, which are computed for the reconstructed and the ground truth attenuation values, as well as MAE in Hounsfield Units. MAE is computed at $2$ mm resolution as well as at the reduced resolution of $4$ mm, where both the ground truth and the reconstruction are binned. The additional MAE computation on downsampled data is designed to provide more insight about HU calibration for radiotherapy applications, since the `ground truth' CT scans were not denoised and thus remain quite noisy as can be seen from the difference maps.

\begin{figure*}[h]
    \centering
    \includegraphics[width=0.95\linewidth]{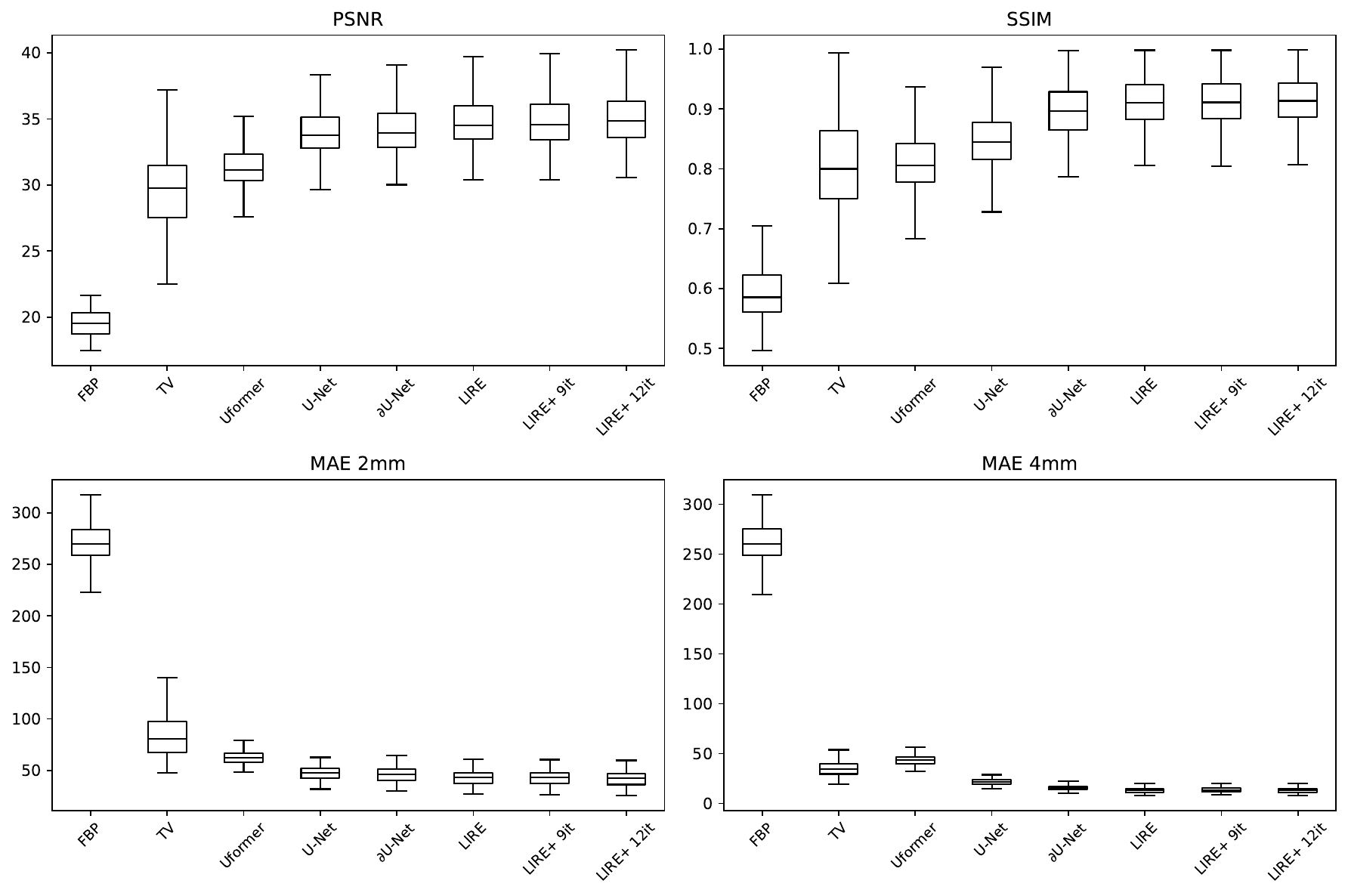}
    \caption{Thorax CT image quality metrics, straight orientation}
    \label{fig:lthorax-str}
\end{figure*}

\begin{figure*}[h]
    \centering
    \includegraphics[width=0.95\linewidth]{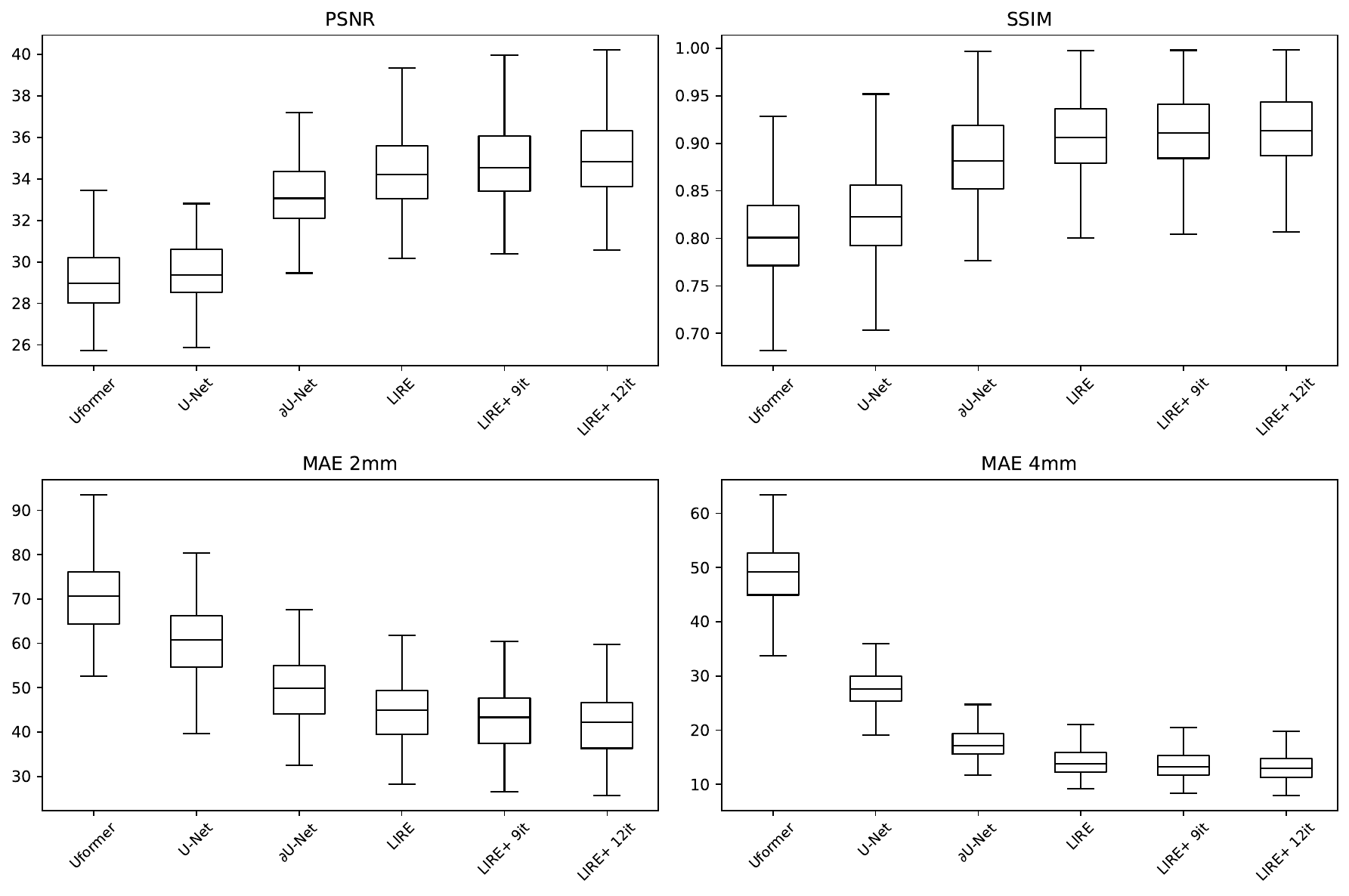}
    \caption{Thorax CT image quality metrics, rotated orientation}
    \label{fig:lthorax-r90}
\end{figure*}

In Tables \ref{tab:str-llung} and \ref{tab:r90-llung} we report these metrics on thorax CT data for straight and rotated patient orientation respectively, and the corresponding box plots are provided in Figures \ref{fig:lthorax-str} and \ref{fig:lthorax-r90}. Since the classical reconstruction methods such as FBP and TV are not trained on specific patient orientation, they are robust to rotations by design and are ommited in the second comparison. On the straight data, we observe that LIRE and LIRE+ noticeably outperform all the baselines. LIRE+ is able to achieve LIRE level of performance using only $9$ iterations out of $12$. Full version of LIRE+ with 12 iterations gives a small performance improvement over LIRE. $\partial$U-net, while being fast, cannot match the reconstruction quality achieved by LIRE/LIRE+, even though it has more parameters. On the rotated data, we note that only LIRE+ is able to maintain the reconstruction quality thanks to the rotationally-equivariant primal cells, while all alternative models suffer from various amounts of performance degradation. It is wothwhile to note that the performance degradation is more pronounced in the learned post-processing baselines (Uformer and U-net), while learned iterative schemes seem to be more robust.

\begin{figure*}[!ht]
    \centering
    \subfloat[]{\includegraphics[width=0.25\linewidth]{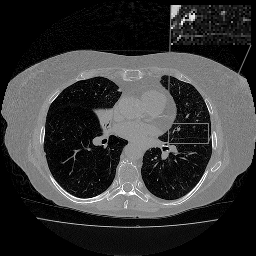}
    }
    \hfil
    \subfloat[]{\includegraphics[ width=0.25\linewidth]{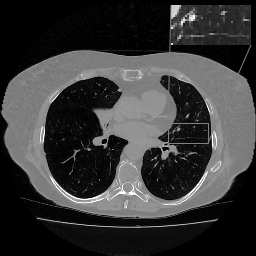}
    }
    \hfil
    \subfloat[]{\includegraphics[ width=0.25\linewidth]{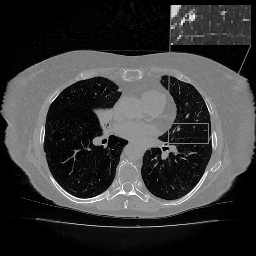}
    }
    \hfil
    \subfloat[]{\includegraphics[ width=0.25\linewidth]{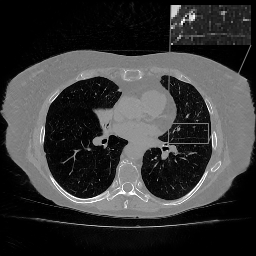}
    }
    \caption{(a) Axial slice of Thorax CT, HU range=(-1000, 800) and (-1350, 150) for ROI, (b) $\partial$U-net, (c) LIRE, and (d) LIRE+ 12 it.}
    \label{fig:axial-large}
\end{figure*}

\begin{figure*}[!ht]
    \centering
    \subfloat[]{\includegraphics[width=0.25\linewidth]{images/cf6c_ax_gt.png}
    }
    \hfil
    \subfloat[]{\includegraphics[ width=0.25\linewidth]{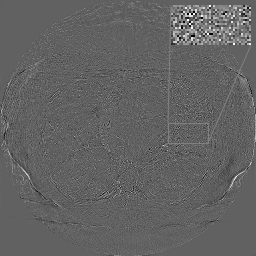}
    }
    \hfil
    \subfloat[]{\includegraphics[ width=0.25\linewidth]{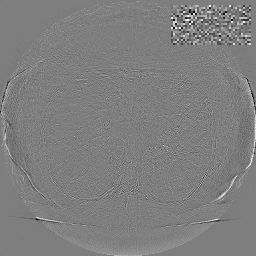}
    }
    \hfil
    \subfloat[]{\includegraphics[ width=0.25\linewidth]{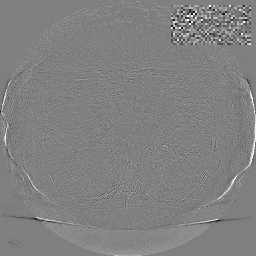}
    }
    \caption{(a) Axial slice of Thorax CT, (b) $\partial$U-net error, HU range=(-1000, 800) and (-200, 200) for ROI, (c) LIRE error, and (d) LIRE+ 12 it. error}
    \label{fig:daxial-large}
  \end{figure*}

\begin{figure*}[!ht]
    \centering
    \subfloat[]{\includegraphics[width=0.25\linewidth]{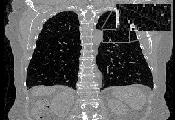}
    }
    \hfil
    \subfloat[]{\includegraphics[ width=0.25\linewidth]{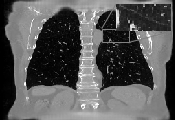}
    }
    \hfil
    \subfloat[]{\includegraphics[ width=0.25\linewidth]{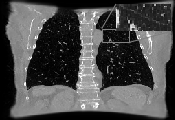}
    }
    \hfil
    \subfloat[]{\includegraphics[ width=0.25\linewidth]{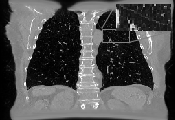}
    }
    \caption{(a) Coronal slice of Thorax CT, HU range=(-1000, 800) and (-1350, 150) for ROI, (b) $\partial$U-net, (c) LIRE, and (d) LIRE+ 12 it.}
    \label{fig:coronal-large}
\end{figure*}

\begin{figure*}[!ht]
    \centering
    \subfloat[]{\includegraphics[width=0.25\linewidth]{images/cf6c_cor_gt.png}
    }
    \hfil
    \subfloat[]{\includegraphics[ width=0.25\linewidth]{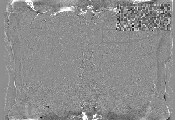}
    }
    \hfil
    \subfloat[]{\includegraphics[ width=0.25\linewidth]{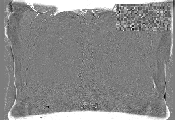}
    }
    \hfil
    \subfloat[]{\includegraphics[ width=0.25\linewidth]{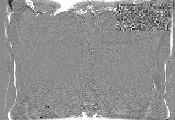}
    }
    \caption{(a) Coronal slice of Thorax CT, (b) $\partial$U-net error, HU range=(-1000, 800) and (-200, 200) for ROI, (c) LIRE error, and (d) LIRE+ 12 it. error}
    \label{fig:dcoronal-large}
  \end{figure*}

Examples of axial image slices of a ground truth image and the corresponding reconstructions with $\partial$U-net, LIRE and LIRE+ are presented in Fig. \ref{fig:axial-large} with the respective difference maps in Fig. \ref{fig:daxial-large}. Coronal view is provided in Fig. \ref{fig:coronal-large} and Fig. \ref{fig:dcoronal-large}. For the image samples the HU range equals (-1000, 800) and (-1350, 150) for the ROI, while for the difference maps HU range equals (-1000, 800) and (-200, 200) for the ROI. From these examples we can see that LIRE+ gives sharper images with better HU calibration, while $\partial$U-net appears to slightly blur lung fissures. The difference maps suggest that particularly for LIRE+ image noise plays a large role in the image quality metrics.

\subsection{Image quality: head \& neck CT}
\begin{table*}[t]
\caption{Test results on Head \& Neck CT without finetuning, mean $\pm$ std.dev. Results on test excluding the outliers given in italic.}
\label{tab:hn-eval}
\centering
\begin{tabular}{|l|l l l l|}
\hline
Method & PSNR & SSIM & MAE 2mm (HU)  & MAE 4mm (HU) \\
  \hline
  TV & \makecell{$37.86 \pm 1.36$ \\ $\mathit{37.92 \pm 1.33}$} & \makecell{$0.94 \pm 0.02$ \\ $\mathit{0.94 \pm 0.02}$} & \makecell{$30.72 \pm 5.63$ \\ $\mathit{30.46 \pm 5.37}$ } & \makecell{$15.48 \pm 3.20$ \\ $\mathit{15.25 \pm 2.74}$} \\
  \cline{2-5}
  $\partial$U-net  & \makecell{$40.24 \pm 1.57$ \\ $\mathit{40.47 \pm 1.18}$} & \makecell{$0.97 \pm 0.01$ \\ $\mathit{0.98 \pm 0.01}$ } & \makecell{$17.81 \pm 4.33$ \\ $\mathit{17.00 \pm 2.11}$ } & \makecell{$9.20 \pm 2.51$ \\ $\mathit{8.74 \pm 1.31}$}\\
  \cline{2-5}
  LIRE & \makecell{$41.21 \pm 1.41$ \\ $\mathit{41.43 \pm 1.22}$ }& \makecell{$0.97 \pm 0.01$ \\ $\mathit{0.97 \pm 0.01}$} & \makecell{$17.75 \pm 2.90$ \\ $\mathit{17.23 \pm 1.87}$} & \makecell{$10.33 \pm 1.57$ \\ $\mathit{10.07 \pm 1.08}$} \\
  \cline{2-5}
  LIRE+ 9 it. & \makecell{$39.66 \pm 9.52$ \\ $\mathit{41.78 \pm 1.66}$ }& \makecell{$0.95 \pm 0.15$ \\ $\mathit{0.98 \pm 0.01}$} & \makecell{$203.32 \pm 1028.00$ \\ $\mathit{15.68 \pm 2.68}$} & \makecell{$245.77 \pm 1288.28$ \\ $\mathit{7.81 \pm 1.27}$} \\
  \cline{2-5}
LIRE+ 12 it. & \makecell{$40.52 \pm 8.66$ \\ $\mathit{42.45 \pm 1.76}$} & \makecell{$0.95 \pm 0.15$ \\ $\mathit{0.98 \pm 0.01}$} & \makecell{$119.90 \pm 548.44$ \\ $\mathit{14.09 \pm 2.82}$ } & \makecell{$139.62 \pm 675.60$ \\ $\mathit{6.96 \pm 1.51}$}\\
\hline
\end{tabular}
\end{table*}

\begin{table*}[h]
\caption{Test results on Head \& Neck CT with finetuning (best result in bold), mean $\pm$ std.dev.}
\label{tab:hn-ft}
\centering
\begin{tabular}{|l|l l l l|}
\hline
Method & PSNR & SSIM & MAE 2mm (HU)  & MAE 4mm (HU) \\
\hline
$\partial$U-Net & $41.57 \pm 1.33$ & $0.98 \pm 0.01$ & $14.72 \pm 2.87$ & $7.31 \pm 1.80$\\
LIRE & $42.76 \pm 1.52$ & $0.99 \pm 0.01$ & $13.02 \pm 2.14$ & $5.77 \pm 1.00$ \\
LIRE+ 9 it. & $43.17 \pm 1.83$ & $0.99 \pm 0.01$ & $12.54 \pm 2.25$ & $5.34 \pm 1.01$ \\
LIRE+ 12 it. & $\mathbf{43.82 \pm 1.93}$ & $\mathbf{0.99 \pm 0.01}$ & $\mathbf{11.50 \pm 2.21}$ & $\mathbf{4.82 \pm 1.02}$ \\
\hline
\end{tabular}
\end{table*}

We perform evaluation on the out-of-distribution HN dataset in two regimes. Firstly, we evaluate the reconstruction performance of LIRE+, LIRE and $\partial$U-net without any finetuning and provide the metrics in Table \ref{tab:hn-eval} in normal font. Inspection of the metrics and the images revealed that there are 4 outlier cases\footnote{The outliers are all very big or tiny patients.} on which LIRE+ performs poorly, while on the majority of cases LIRE+ is comparable to LIRE. Reconstruction metrics on the HN set with the outlier cases excluded are provided in Table \ref{tab:hn-eval} in italic font, indicating that LIRE+ might actually outperform LIRE on a majority of HN cases. For comparison, iterative reconstruction baseline is provided as well.

\begin{figure*}[!ht]
    \centering
    \subfloat[]{\includegraphics[width=0.25\linewidth]{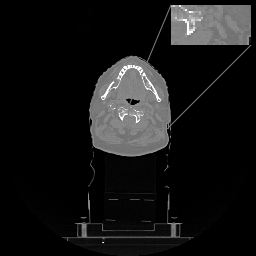}
    }
    \hfil
    \subfloat[]{\includegraphics[ width=0.25\linewidth]{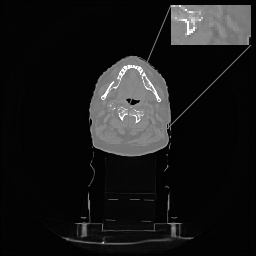}
    }
    \hfil
    \subfloat[]{\includegraphics[ width=0.25\linewidth]{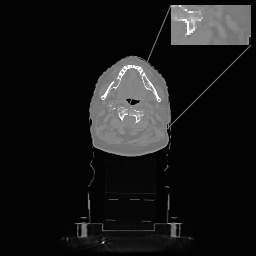}
    }
    \hfil
    \subfloat[]{\includegraphics[ width=0.25\linewidth]{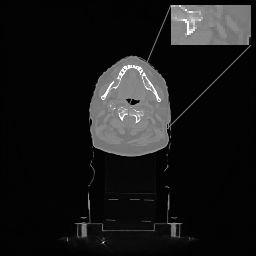}
    }
    \caption{(a) Axial slice of HN CT, HU range=(-1000, 1000), (b) $\partial$U-net, (c) LIRE, and (d) LIRE+ 12 it.}
    \label{fig:axial-large-hn}
  \end{figure*}

Secondly, to further investigate the generalization behaviour of LIRE+, we report the results of the finetuning experiment in Table \ref{tab:hn-ft}. Interestingly, after identical finetuning, both full LIRE+ and LIRE+ with only 9 iterations demonstrate substantially better performance compared to LIRE and $\partial$U-net. The better generalization of LIRE+ after finetuning on a limited amount of data is in agreement with the lower parameter count of the new model. Axial image slices from finetuned HN models are provided in Figure \ref{fig:axial-large-hn}, where HU range is set to (-1000, 1000).

\section{Discussion}
\label{s.discussion}
We have presented LIRE+, a fast, compact and memory-efficient multiscale equivariant learned iterative scheme for CBCT reconstruction. Compared to LIRE\cite{lire2023}, LIRE+
has substantially lower parameter count, similar inference time for better image quality or twice faster inference for similar image quality, and enjoys additional robustness to patient
orientation, which we achieve by using rotationally-equivairant primal blocks. It is noteworthy that LIRE+ is the smallest deep learning reconstruction model in our comparison, but it still gives the best image quality. On the out-of-distribution head \& neck dataset, we have observed that LIRE+ is generally comparable to LIRE but has some outlier cases, however, after identical finetuning on a limited amount of head \& neck data LIRE+ strongly outperforms LIRE and $\partial$U-net.

The authors would like to acknowledge the Research High Performance Computing (RHPC) facility of the Netherlands Cancer Institute (NKI).


\end{document}